\documentclass[sigconf]{acmart}

\usepackage{amsmath,amsfonts}
\usepackage{algorithm,algorithmicx}
\usepackage{algpseudocode}
\usepackage{marvosym}
\usepackage{threeparttable}
\usepackage{graphicx}
\usepackage{tabularx}
\usepackage{caption}
\usepackage{multirow}
\usepackage{mathrsfs}
\usepackage{textcomp}
\usepackage{xcolor}
\usepackage{verbatim}
\usepackage{float} 
\usepackage{grffile}
\usepackage{makecell}
\usepackage{pifont}
\usepackage{tikz}
\usepackage{booktabs}
\usepackage{amsthm}
\usepackage{balance}
\usepackage{subfigure}
\usepackage{array}
\usepackage{pdfpages}
\usepackage{stfloats}
\usepackage{url}
\usepackage{stmaryrd}
\usepackage{colortbl}

\renewcommand\footnotetextcopyrightpermission[1]{} 

\AtBeginDocument{%
  }

\makeatletter
\newenvironment{breakablealgorithm}
{
		\begin{center}
			\refstepcounter{algorithm}
			\hrule height.8pt depth0pt \kern2pt
			\renewcommand{\caption}[2][\relax]{
				{\raggedright\textbf{\ALG@name~\thealgorithm} ##2\par}%
				\ifx\relax##1\relax 
				\addcontentsline{loa}{algorithm}{\protect\numberline{\thealgorithm}##2}%
				\else 
				\addcontentsline{loa}{algorithm}{\protect\numberline{\thealgorithm}##1}%
				\fi
				\kern2pt\hrule\kern2pt
			}
		}{
		\kern2pt\hrule\relax
	\end{center}
}

\newcommand*\halfcirc[1][1ex]{%
	\begin{tikzpicture}
	\draw[fill] (0,0)-- (90:#1) arc (90:270:#1) -- cycle ;
	\draw (0,0) circle (#1);
	\end{tikzpicture}}
\newcommand*\fullcirc[1][1ex]{\tikz\fill (0,0) circle (#1);} 

\newtheorem{definition}{Definition}

\definecolor{lightgray}{gray}{0.9}

\begin{document}

\title{PraxiMLP: A Threshold-based Framework for Efficient Three-Party MLP with Practical Security}



\author{
Tianle Tao\textsuperscript{1}, 
Shizhao Peng\textsuperscript{1}, 
Haogang Zhu\textsuperscript{1,2,3,\Letter}
}

\affiliation{
 \institution{\textsuperscript{1}State Key Laboratory of Complex \& Critical Software Environment, Beihang University, Beijing, China   \\
 \textsuperscript{2}Zhongguancun Laboratory, Beijing, China \\
 \textsuperscript{3}Hangzhou International Innovation Institute, Beihang University, Hangzhou, China
 }
 \country{}
}
\affiliation{
  \institution{\{taotianle,\hspace{0.2em}by1806167,\hspace{0.2em}haogangzhu\}@buaa.edu.cn}
  \country{}
}


\begin{abstract}
Efficiency and communication cost remain critical bottlenecks 
for practical Privacy-Preserving Machine Learning (PPML). 
Most existing frameworks rely on fixed-point arithmetic for strong security, 
which introduces significant precision loss and requires expensive 
cross-domain conversions (e.g., Arithmetic-to-Boolean) 
for non-linear operations. 
To address this, we propose PraxiMLP, a highly efficient three-party 
MLP framework grounded in practical security. 
The core of our work is a pair of novel additive-to-multiplicative 
conversion protocols that operate entirely within the arithmetic domain, 
thus avoiding expensive cross-domain conversions. 
By natively supporting floating-point numbers, PraxiMLP precisely 
handles non-linear functions, dramatically improving both efficiency 
and precision. Experimental results confirm that, 
compared to mainstream PPML frameworks, PraxiMLP delivers an 
average 8 orders of magnitude precision improvement on basic protocols and 
a 5$\times$ average model training speedup in a WAN environment.
\end{abstract}

\keywords{Privacy-preserving Machine Learning, Practical Security, Neural Network, Threshold}

\maketitle

\section{Introduction}


The rapid advancement of Artificial Intelligence (AI) has injected new momentum 
into various industries, significantly boosting productivity and underscoring the immense value of data. 
However, training high-performance models is heavily dependent on high-quality data, 
creating a fundamental tension with the growing global concern for data privacy 
(e.g., the GDPR\cite{cite-GDPR} enacted by the European Union).
Consequently, securely leveraging private data for 
machine learning has emerged as a pivotal research topic. Privacy-Preserving Machine 
Learning (PPML) offers a promising solution by integrating machine learning with 
Privacy-Enhancing Technologies, such as Secure Multi-Party Computation (MPC)\cite{cite-MPC}, 
Homomorphic Encryption (HE)\cite{cite-HE}, Federated Learning (FL)\cite{cite-FL}, 
and Differential Privacy (DP)\cite{cite-DP}.


Machine learning data is predominantly real-valued,
and models rely heavily on non-linear functions (e.g., Softmax).
However, in pursuit of rigorous security, 
most PPML frameworks are constrained to fixed-point arithmetic 
over finite fields or rings\cite{cite-MPC-NonLinear}.
This necessitates approximating non-linearities with linear and 
comparison operations, leading to inherent losses in efficiency and precision.
Such constraints are often at odds with practical applications, 
which demand a pragmatic trade-off among security, efficiency, and accuracy.
For example, when collaboratively training a user preference model on massive datasets,
e-commerce platforms might be willing to relax the privacy guarantees on 
macro-level statistics (e.g., the general distribution of purchase rates) in exchange 
for the the efficiency and accuracy required for a commercially viable model.
This highlights the significant research value of PPML approaches designed for 
practical security.


Inspired by the data disguising (DD) technique and heuristic practical security model 
from Du et al.\cite{cite-Du}, several frameworks\cite{cite-EVA-S3PC, cite-EVA-S2PLoR} have been 
developed for practical secure linear and logistic regression.
However, these frameworks suffer from key limitations: 
they rely on a semi-honest commodity server for assistance 
and lack support for neural networks. As a cornerstone of deep learning, 
neural networks enable powerful data analysis across critical domains like 
healthcare, transportation, and finance. To address this gap, this paper
proposes PraxiMLP, a 2-out-of-3 threshold-based three-party framework for 
Multi-Layer Perceptrons (MLPs)\cite{cite-MLP} with practical security. 
The primary contributions of this work are as follows:







\begin{itemize}




    \item The design of novel conversion protocols between 
    additive and multiplicative sharing that operate purely within 
    the arithmetic domain to support non-linear functions.

    \item The introduction of the PraxiMLP framework, a secure 
    three-party MLP built on 2-out-of-3 threshold protocols 
    (e.g., MatMul, ReLU, Softmax), accompanied by a 
    theoretical analysis of complexity and practical security.

    \item A C++ implementation and comprehensive experimental 
    benchmark of the PraxiMLP framework against mainstream counterparts. 
    The results demonstrate an average $10^8\times$ precision gain on 
    basic protocols and a $5\times$ average speedup for model training 
    under WAN conditions.

\end{itemize}

\section{Related Work}



A representative framework in PPML is SecureML\cite{cite-SecureML}, which
integrates multiple MPC techniques like Secret Sharing (SS)\cite{cite-SS}, 
Oblivious Transfer (OT)\cite{cite-OT}, and Garbled Circuits (GC)\cite{cite-GC} 
for secure two-party MLP models. ABY3\cite{cite-ABY3} introduced protocols for 
efficient conversions among arithmetic, Boolean, and Yao's circuits.
In conjunction with 2-out-of-3 replicated secret sharing (RSS)\cite{cite-RSS}, 
this enables the efficient implementation of secure 
three-party MLP and Convolutional Neural Network (CNN)\cite{cite-CNN} models.
SecureNN\cite{cite-SecureNN} presented optimized, GC-free protocols for 
nonlinear functions (e.g., ReLU) under the malicious model.
Modern platforms, such as MP-SPDZ\cite{cite-MP-SPDZ}, CrypTen\cite{cite-CrypTen}, 
and SecretFlow\cite{cite-SecretFlow}, integrate a wide array of advanced MPC protocols, 
providing high-level APIs and strong usability to facilitate the 
efficient training of neural networks.

Du et al.\cite{cite-Du} pioneered the use of the data disguising technique 
to design a practical secure machine learning framework for linear regression 
and classification. Building upon this work, EVA-S3PC\cite{cite-EVA-S3PC} proposed 
a suite of secure three-party matrix operation protocols and 
combined them with Monte Carlo methods to construct a verifiable 
secure three-party linear regression model. Separately, EVA-S2PLoR\cite{cite-EVA-S2PLoR} 
designed and implemented a secure two-party logistic regression model based on a 
verifiable Hadamard product protocol. Table \ref{tab: PPML Comparison} compares several representative PPML frameworks.

\begin{table}[htbp]
  \centering
  \caption{Comparison of Several PPML Frameworks} 
  \label{tab: PPML Comparison}
    
  \resizebox{\linewidth}{!}{
    \begin{threeparttable} 
      
      \begin{tabular}{cccccc}
      \toprule 
      \textbf{Type} & \textbf{Framework} & \textbf{Dev. Language} & \textbf{Security} & \textbf{ML Supported} & \textbf{Complexity}\\
      \midrule
      \multirow{6}[1]{*}{\textbf{MPC-based}} 
      & SecureML\cite{cite-SecureML} & C++ & Semi-honest & \fullcirc & High \\
      & ABY3\cite{cite-ABY3} & C++ & Semi-honest & \fullcirc & Medium \\
      & SecureNN\cite{cite-SecureNN} & C++ & Malicious & \fullcirc & High \\
      & MP-SPDZ\cite{cite-MP-SPDZ} & C++ & Malicious & \fullcirc & Medium \\
      & CrypTen\cite{cite-CrypTen} & Python & Semi-honest & \fullcirc & Medium \\
      & SecretFlow\cite{cite-SecretFlow} & C++ & Malicious & \fullcirc & Medium \\
      \midrule
      \multirow{4}[1]{*}{\textbf{DD-based}}
      & Du et al.\cite{cite-Du} & / & Semi-honest & \halfcirc & Low \\
      & EVA-S3PC\cite{cite-EVA-S3PC} & Python & Semi-honest & \halfcirc & Low \\
      & EVA-S2PLoR\cite{cite-EVA-S2PLoR} & Python & Semi-honest & \halfcirc & Low \\
      & PraxiMLP & C++ & Semi-honest & \fullcirc & Low \\
      \bottomrule
      \end{tabular}
      
      \begin{tablenotes}
          \small 
          \item \textbf{Note:} In this table, \fullcirc\: supports neural networks, 
          while \halfcirc\: supports only linear/logistic regression.
      \end{tablenotes}

    \end{threeparttable} 
  } 
\end{table}%

\section{Preliminary}


This section introduces the necessary preliminaries.

\subsection{Practical Security Model}


The definitions for the practical security model 
in this paper are presented below.

\begin{definition}[Semi-honest Adversary Model \cite{cite-Semi-honest}]\label{def1}

In the semi-honest adversary model, all participating parties are 
assumed to adhere to the protocol specification, 
but they may attempt to infer private information 
from their own view of the protocol execution.

\end{definition}

\begin{definition}[Semi-honest Practical Security Model\cite{cite-EVA-S2PLoR}]

A protocol achieves practical security in the semi-honest model 
if any inferred range for a participant's private data is still 
infinite or sufficiently large to resist brute-force computation,
and thus remains acceptable in the given context.
\end{definition}

\subsection{Practical Replicated Secret Sharing}


Our Practical Replicated Secret Sharing (PRSS) scheme, 
similar to standard RSS\cite{cite-RSS}, is a 2-out-of-3 
threshold protocol among three parties $(P_0, P_1, P_2)$. 
For a secret $x\in \mathbb{R}$, an 
additive sharing, denoted by $\llbracket x \rrbracket$, is generated by 
randomly choosing shares $x_0, x_1, x_2 \in \mathbb{R}$ satisfying 
$x = x_0 + x_1 + x_2$. The shares are distributed such that $P_0$ holds 
$(x_0, x_1)$, $P_1$ holds $(x_1, x_2)$, and $P_2$ holds $(x_2, x_0)$. 
We also define a multiplicative variant, denoted by 
$\llbracket x \rrbracket ^{M}$, where the shares satisfy 
$x = x_0 \cdot x_1 \cdot x_2$. As any pair of parties holds all 
three unique shares, both schemes provide a 2-out-of-3 threshold.


Additionally, inspired by \cite{cite-RSS}, we employ a pseudorandom generator 
(PRG) to generate correlated randomness. Specifically, during a one-time setup phase, 
each party $P_i$ generates a random seed $s_i$ and sends it to the next party in a cycle, 
resulting in the distribution: $P_0$ holds $(s_0, s_1)$, $P_1$ holds $(s_1, s_2)$, 
and $P_2$ holds $(s_2, s_0)$. Any pair of parties $(P_{i-1}, P_{i})$ 
can then derive a common random value $r_i$ by applying the PRG to their shared seed $s_i$.
This operation is denoted generally as $r_i = \text{PRG}(s_i, P_{i-1}, P_i)$.

\subsection{Notation}
Throughout this paper, matrix multiplication is denoted by $x \times y$, 
and the Hadamard product by $x \odot y$. 
Unless specified otherwise, all other scala operations, 
such as $x / y$, $x + 1$, and $e^x$, are applied to matrices element-wise. 
For simplicity of notation, subscripts index the parties  
($P_i$ for $i \in \{0, 1, 2\}$), and all subscript arithmetic 
is performed in the ring $\mathbb{Z}_3$ (i.e., modulo 3).

\section{The PraxiMLP Framework}


This section details the architecture of the PraxiMLP framework and the specific 
procedures of its internal protocols.



\subsection{Framework}

As a foundational neural network architecture, 
the MLP is typically composed of alternating linear layers and 
non-linear activation functions. This paper proposes PraxiMLP, 
a secure three-party MLP framework for 
secure Machine Learning as a Service (MLaaS)\cite{cite-MLaaS}. 
The framework utilizes the ReLU function for its hidden layer activations, 
Softmax for the output layer, and employs the cross-entropy loss function 
during training.

\begin{figure}[ht]
  \centering
  \includegraphics[width=1.0\hsize]{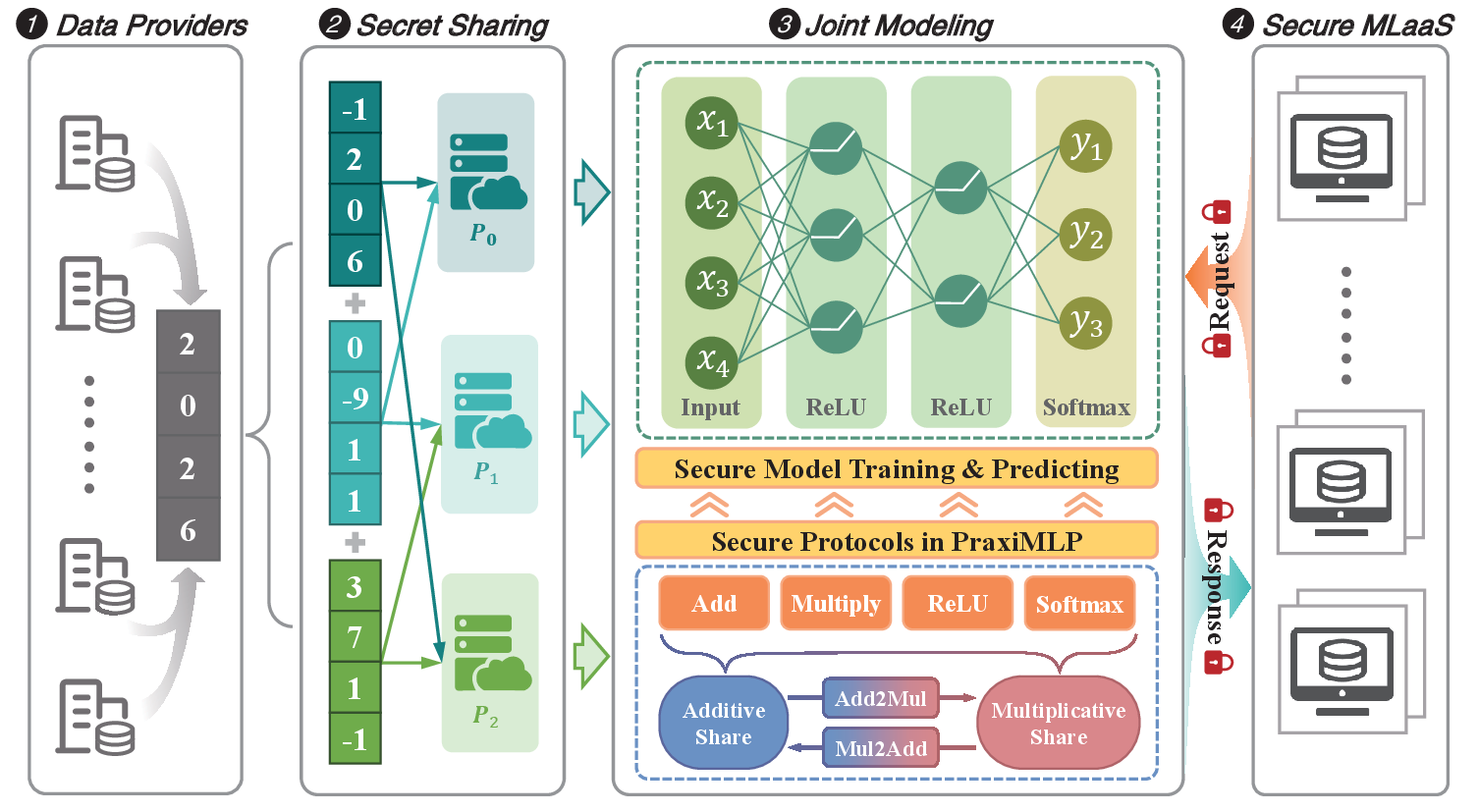}
  \caption{The PraxiMLP Framework}
  \label{fig:PraxiMLP-Framework}
\end{figure}

The PraxiMLP architecture consists of three computation servers operating 
under a 2-out-of-3 threshold scheme, thus tolerating 
the corruption of a single server. The workflow begins with 
data providers distributing shares of their private data to 
the servers via the PRSS scheme. 
Subsequently, the servers execute a suite of secure computation protocols to 
perform both joint model training and subsequent secure user inference. 
After the model is trained, users can perform secure, interactive inference 
with the model deployed on the servers. A schematic diagram of the 
PraxiMLP framework is illustrated in Figure \ref{fig:PraxiMLP-Framework}.

Training and inference in PraxiMLP primarily consist of linear operations, 
activation functions (ReLU, Softmax), and gradient computations. 
For a Softmax output layer, 
the gradient of the cross-entropy loss has the simple analytical form: 
$G = \hat{Y} - Y$, where $\hat{Y}$ is the model prediction and $Y$ represents 
the ground-truth label. Gradient computations for the hidden layers rely only on 
basic matrix operations and the ReLU derivative. Therefore, the framework's entire 
secure computation requirements reduce to a core set of secure protocols for 
linear operations (matrix multiplication, addition, Hadamard product) and 
activation functions (ReLU and its derivative, Softmax), 
which are detailed in the subsequent sections.

\subsection{Secure Protocols}




\subsubsection{Sharing and Reconstruction}
During the secret sharing process, a sharing of zero is generated, where each party 
$P_i$ obtains a share $\alpha_i$ such that $\sum_{i=0}^{2}{\alpha_i}=0$. 
This is achieved by first leveraging the PRG to generate a random sharing $\llbracket r \rrbracket$, 
where $r_i = \text{PRG}(s_{i}, P_{i-1}, P_{i})$. Each party $P_i$ then locally computes 
its share as $\alpha_i = r_i - r_{i+1}$. This entire process is 
denoted as $(\alpha_0, \alpha_1, \alpha_2) = \text{Shr}(0)$.

When party $P_0$ shares its private value $x$, a sharing of zero 
$(\alpha_0, \alpha_1, \alpha_2)$ is generated first. Then $\llbracket x \rrbracket$ 
is constructed with the shares $(x_0, x_1, x_2) = (x+\alpha_0, \alpha_1, \alpha_2)$. 
Each party $P_i$ sends its corresponding share $x_i$ to party $P_{i-1}$. 
This procedure is abbreviated as $\llbracket x \rrbracket = \text{Shr}_{0}(x)$. 
The sharing process for other parties is analogous.

To reconstruct the secret $x$, each party $P_i$ sends its share $x_i$ to 
party $P_{i+1}$. The secret can then be recovered by locally computing 
the sum $x = x_0 + x_1 + x_2$. This procedure is denoted as 
$x = \text{Rec}(\llbracket x \rrbracket)$. 
The reconstruction can also be restricted to specific parties. 
For example, revealing the secret only to $P_0$, 
denoted $x = \text{Rec}_{0}(\llbracket x \rrbracket)$, 
is achieved by having $P_2$ send its share $x_2$ to $P_0$.
Evidently, the procedures described above are applicable to both scalars and matrices.

\subsubsection{Addition and Multiplication}
Computing the sum $\llbracket z \rrbracket = \llbracket x \rrbracket + 
\llbracket y \rrbracket$ is straightforward, as it is a purely local operation. 
Each party $P_i$ simply needs to compute the sum of its held shares: 
$(z_i, z_{i+1}) = (x_i + y_{i}, x_{i+1} + y_{i+1})$.

The computation of the product 
$\llbracket z \rrbracket = \llbracket x \rrbracket \cdot \llbracket y \rrbracket$ 
is more involved. 
\begin{align}
\llbracket x \rrbracket \cdot \llbracket y \rrbracket 
&= (x_0+x_1+x_2) \cdot (y_0 + y_1 + y_2) \nonumber \\ 
&= x_0\cdot (y_0 + y_1) + x_1 \cdot y_0 \nonumber \\
&+ x_1 \cdot (y_1 + y_2) + x_2 \cdot y_1 \nonumber \\
&+ x_2 \cdot (y_2 + y_0) + x_0 \cdot y_2 
\end{align}

Each party $P_i$ first generates a sharing of zero $(\alpha_0, \alpha_1, \alpha_2)$, 
then locally computes a partial product 
$z_i = x_i \cdot (y_i + y_{i+1}) + x_{i+1} \cdot y_i + \alpha_i$, 
and finally sends $z_i$ to $P_{i-1}$.
The above procedure applies to both scalars and matrices, 
with scalar multiplication, matrix multiplication, 
and the Hadamard product denoted by 
$\cdot$, $\times$, and $\odot$, respectively. 
Additionally, the multiplication of a shared value by 
a plaintext (e.g., $\llbracket z \rrbracket = \llbracket x \rrbracket \cdot y$) 
is a straightforward local computation akin to addition. 
The details are therefore omitted for brevity.



\subsubsection{Share Conversion}
The PraxiMLP framework requires conversions between additive and 
multiplicative sharings, denoted as 
$\llbracket x \rrbracket^{M} = \text{Add2Mul}(\llbracket x \rrbracket)$ and 
$\llbracket x \rrbracket = \text{Mul2Add}(\llbracket x \rrbracket^{M})$, respectively.
We assume, without loss of generality, that these protocols 
operate on matrices, where the multiplication in a 
multiplicative sharing is the Hadamard product.

Let the output of the $\text{Add2Mul}(\llbracket x \rrbracket)$ be 
$\llbracket x \rrbracket^{M} = (\hat{x}_0, \hat{x}_1, \hat{x}_2)$. 
The protocol begins by generating the first two shares as random matrices 
with non-zero elements: $\hat{x}_0 = \text{PRG}(s_0, P_2, P_0)$, 
and $\hat{x}_1 = \text{PRG}(s_1, P_0, P_1)$.
The third share, $\hat{x}_2$, must then be computed to satisfy the equation 
$\hat{x}_2 = x / (\hat{x}_0 \odot \hat{x}_1)$
To achieve this securely, $P_0$ (the only party holding both $\hat{x}_0$ and $\hat{x}_1$) 
locally computes $t = 1 / (\hat{x}_0 \odot \hat{x}_1)$, and then shares this value: 
$\llbracket t \rrbracket = \text{Shr}_0(t)$.
This allows all parties to compute 
$\llbracket \hat{x}_2 \rrbracket = \llbracket x \rrbracket \odot 
\llbracket t \rrbracket$ via one secure Hadamard product.
Finally, the resulting share $\hat{x}_2$ is reconstructed to parties $P_1$ and $P_2$. 
The detailed procedure of this protocol is formalized in Algorithm \ref{alg:Add2Mul}.

\begin{breakablealgorithm}
    \caption{Add2Mul}
    \label{alg:Add2Mul}
    \begin{algorithmic}[1] 
        \Require $\llbracket x \rrbracket=(x_0,x_1,x_2)$ where $x_0,x_1,x_2 \in \mathbb{R}^{n\times m}$
        \Ensure $\llbracket x \rrbracket^M$
        \State $\hat{x}_0 = \text{PRG}(s_0, P_2, P_0)$ \Comment{$\hat{x}_0\in \mathbb{R}^{n\times m}$}
        \State $\hat{x}_1 = \text{PRG}(s_1, P_0, P_1)$ \Comment{$\hat{x}_1\in \mathbb{R}^{n\times m}$}
        \State $P_0$ computes $t = 1 / (\hat{x}_0 \odot \hat{x}_1)$
        \State $\llbracket t \rrbracket=\text{Shr}_0(t)$
        \State $\llbracket \hat{x}_2 \rrbracket=\llbracket x \rrbracket \odot \llbracket t \rrbracket$
        \State $\hat{x}_2 = \text{Rec}_{1,2}(\llbracket \hat{x}_2 \rrbracket)$
        \State \Return $\llbracket x \rrbracket^M = (\hat{x}_0, \hat{x}_1, \hat{x}_2)$
    \end{algorithmic}
\end{breakablealgorithm}


Steps 5 and 6 in Algorithm \ref{alg:Add2Mul} can be fused. 
After the local computations of Step 5, the intermediate results can be 
directly sent to $P_1$ and $P_2$ for the reconstruction of $\hat{x}_2$, 
reducing the protocol to two communication rounds.
Correctness is confirmed by the following calculation: 
$\hat{x}_0 \odot \hat{x}_1 \odot \hat{x}_2 = \hat{x}_0 \odot \hat{x}_1 \odot 
(x_0 + x_1 + x_2) / (\hat{x}_0 \odot \hat{x}_1) = x$.


The $\text{Mul2Add}(\llbracket x \rrbracket^M)$ protocol, 
which converts a multiplicative sharing $\llbracket x \rrbracket^M = 
(\hat{x}_0, \hat{x}_1, \hat{x}_2)$ to an additive one, 
mirrors the secure multiplication logic. It requires two parallel 
secret-sharing inputs for a final secure multiplication round:  
$P_0$ locally computes $t = \hat{x}_0 \odot \hat{x}_1$ and 
initiates $\llbracket t \rrbracket = \text{Shr}_0(t)$, 
while $P_2$ initiates $\llbracket \hat{x}_2 \rrbracket = \text{Shr}_2(\hat{x}_2)$.
The final additive sharing is then computed as $\llbracket x \rrbracket = 
\llbracket t \rrbracket \odot \llbracket \hat{x}_2 \rrbracket$. 
The proof of correctness is inherent in the computational procedure itself. 
The full protocol is specified in Algorithm \ref{alg:Mul2Add}.

\begin{breakablealgorithm}
    \caption{Mul2Add}
    \label{alg:Mul2Add}
    \begin{algorithmic}[1] 
        \Require $\llbracket x \rrbracket^M = (\hat{x}_0, \hat{x}_1, \hat{x}_2)$ 
        where $\hat{x}_0, \hat{x}_1, \hat{x}_2 \in \mathbb{R}^{n\times m}$
        \Ensure $\llbracket x \rrbracket$
        \State $P_0$ computes $t = \hat{x}_0 \odot \hat{x}_1$
        \State $\llbracket t \rrbracket=\text{Shr}_0(t)$
        \State $\llbracket \hat{x}_2 \rrbracket=\text{Shr}_2(\hat{x}_2)$
        \State $\llbracket x \rrbracket = \llbracket t \rrbracket \odot 
        \llbracket \hat{x}_2 \rrbracket$
        \State \Return $\llbracket x \rrbracket$
    \end{algorithmic}
\end{breakablealgorithm}




\subsubsection{ReLU Function}
The computation of $\text{ReLU}(x) = \max(0, x)$ is based on 
its derivative, $\text{ReLU}'(x)$, which satisfies the property $\text{ReLU}(x) = \text{ReLU}'(x) \cdot x$. 
The derivative takes the value of 1 for $x \ge 0$ and 0 otherwise.
By defining the $\text{Sign}(x)$ to be 1 for $x \ge 0$ and -1 otherwise, 
we arrive at the key identity $\text{ReLU}'(x) = (\text{Sign}(x)+1)/2$.

Our protocol first focuses on computing $\text{ReLU}'(\llbracket x \rrbracket)$. 
The parties jointly compute a multiplicative sharing 
$\llbracket x \rrbracket^{M} = (\hat{x}_0, \hat{x}_1, \hat{x}_2)$ by 
invoking $\text{Add2Mul}(\llbracket x \rrbracket)$.
Then each party $P_i$ locally computes the sign of 
its respective shares to form a new multiplicative sharing 
$\llbracket \hat{y} \rrbracket^{M} = (\hat{y}_0, \hat{y}_1, \hat{y}_2)$, 
where $\hat{y}_i = \text{Sign}(\hat{x}_i)$.
The parties then compute the corresponding additive sharing 
$\llbracket \hat{y} \rrbracket = \text{Mul2Add}(\llbracket \hat{y} \rrbracket^M)$, and 
finally compute $\text{ReLU}'(\llbracket x \rrbracket) = \llbracket y \rrbracket = 
(\llbracket \hat{y} \rrbracket + 1) / 2$ locally.

The correctness of this procedure can be verified as follows. 
Since the value of $\text{ReLU}'(0)$ (whether 0 or 1) does not affect the final result of 
$\text{ReLU}(0)=0$, we can disregard the special case where $x=0$.
This allows us to use the property that the product of signs is 
the sign of the product: $\hat{y} = \text{Sign}(\hat{x}_0) \cdot \text{Sign}(\hat{x}_1) 
\cdot \text{Sign}(\hat{x}_2) = \text{Sign}(\hat{x}_0 \cdot \hat{x}_1 \cdot \hat{x}_2) = \text{Sign}(x)$. 
Therefore, the reconstructed plaintext value 
$y = (\hat{y} + 1) / 2$ correctly evaluates to 
$(\text{Sign}(x)+1)/2 = \text{ReLU}'(x)$, 
confirming the correctness of the computation.

\begin{breakablealgorithm}
    \caption{ReLU}
    \label{alg:ReLU}
    \begin{algorithmic}[1] 
        \Require $\llbracket x \rrbracket=(x_0,x_1,x_2)$ 
        where $x_0,x_1,x_2 \in \mathbb{R}^{n\times m}$
        \Ensure $\llbracket y \rrbracket = \text{ReLU}'(\llbracket x \rrbracket)$ and
        $\llbracket z \rrbracket = \text{ReLU}(\llbracket x \rrbracket)$
        \State $\llbracket x \rrbracket^M = \text{Add2Mul}(\llbracket x \rrbracket)$
        \State $\llbracket \hat{y} \rrbracket^M = \text{Sign}(\llbracket x \rrbracket^M)$
        \State $\llbracket \hat{y} \rrbracket = \text{Mul2Add}(\llbracket \hat{y} \rrbracket^M)$
        \State $\llbracket y \rrbracket = (\llbracket \hat{y} \rrbracket + 1) / 2$
        \State $\llbracket z \rrbracket = \llbracket y \rrbracket \odot \llbracket x \rrbracket$
        \State \Return $\llbracket y \rrbracket$ and $\llbracket z \rrbracket$
    \end{algorithmic}
\end{breakablealgorithm}

With the derivative sharing $\llbracket y \rrbracket$ correctly computed, 
the parties obtain the final result via one secure multiplication: 
$\text{ReLU}(\llbracket x \rrbracket) = \text{ReLU}'(\llbracket x \rrbracket) \cdot 
\llbracket x \rrbracket = \llbracket y \rrbracket \cdot \llbracket x \rrbracket$. 
Evidently, this computational procedure can be directly extended to element-wise 
matrix operations. The detailed protocol is presented in Algorithm \ref{alg:ReLU}.



\subsubsection{Softmax Function}
In this paper,the Softmax function is assumed to process all 
row vectors of an input matrix in parallel.
For a single row vector $v = (v_1, v_2, \cdots, v_n)$, the Softmax is defined as 
$\text{Softmax}(v) = (e^{v_1}, e^{v_2}, \cdots, e^{v_n}) / \sum_{i=1}^{n} e^{v_i}$. 
Consequently, for a matrix $x$, this can be expressed as 
$\text{Softmax}(x) = e^{x} / \text{RowSum}(e^x)$, where $\text{RowSum}$ sums each row of $x$ 
and broadcasts the resulting column vector back to the original matrix dimensions.

To compute $\text{Softmax}(\llbracket x \rrbracket)$ on a shared matrix 
$\llbracket x \rrbracket = (x_0, x_1, x_2)$, the protocol proceeds as follows.
First, each party locally exponentiates its share, resulting in a multiplicative 
sharing $\llbracket \hat{x} \rrbracket^M = (e^{x_0}, e^{x_1}, e^{x_2})$. 
This is then converted to an additive sharing via 
$\llbracket \hat{x} \rrbracket = \text{Mul2Add}(\llbracket \hat{x} \rrbracket^M)$. 
The parties then locally compute 
$\llbracket t \rrbracket = \text{RowSum}(\llbracket \hat{x} \rrbracket)$, which is 
subsequently converted back to a multiplicative sharing 
$\llbracket t \rrbracket^M = \text{Add2Mul}(\llbracket t \rrbracket)$.
This allows for element-wise division on the 
multiplicative shares: $\llbracket y \rrbracket^M = 
\llbracket \hat{x} \rrbracket^M / \llbracket t \rrbracket^M$ $= 
(e^{x_0} / t_0, e^{x_1} / t_1, e^{x_2} / t_2)$ locally, which is 
finally converted to the result 
$\llbracket y \rrbracket = \text{Mul2Add}(\llbracket y \rrbracket^M)$. 
Correctness stems from the plaintext reconstruction: 
$y = (e^{x_0} \odot e^{x_1} \odot e^{x_2}) / (t_0 \odot t_1 \odot t_2) 
= e^x / \text{RowSum}(e^x)$. 
The detailed protocol is formalized in Algorithm \ref{alg:Softmax}.

\begin{breakablealgorithm}
    \caption{Softmax}
    \label{alg:Softmax}
    \begin{algorithmic}[1] 
        \Require $\llbracket x \rrbracket=(x_0,x_1,x_2)$ 
        where $x_0,x_1,x_2 \in \mathbb{R}^{n\times m}$
        \Ensure $\llbracket y \rrbracket = \text{Softmax}(\llbracket x \rrbracket)$
        \State $\llbracket \hat{x} \rrbracket^M = (e^{x_0}, e^{x_1}, e^{x_2})$
        \State $\llbracket \hat{x} \rrbracket = \text{Mul2Add}(\llbracket \hat{x} \rrbracket^M)$
        \State $\llbracket t \rrbracket = \text{RowSum}(\llbracket \hat{x} \rrbracket)$
        \State $\llbracket t \rrbracket^M = \text{Add2Mul}(\llbracket t \rrbracket)$
        \State $\llbracket y \rrbracket^M = 
        \llbracket \hat{x} \rrbracket^M / \llbracket t \rrbracket^M $
        \State $\llbracket y \rrbracket = \text{Mul2Add}(\llbracket y \rrbracket^M)$

        \State \Return $\llbracket y \rrbracket$
    \end{algorithmic}
\end{breakablealgorithm}


\section{Theoretical Analysis}

This section presents the complexity and security analysis of the proposed protocols.


\subsection{Complexity Analysis}

Computationally, our secure protocols incur a constant-factor 
overhead compared to their plaintext counterparts, 
and thus share the same asymptotic complexity. 
The communication complexity is evaluated based on two metrics: 
the total data transmitted by all parties (communication cost) and 
the number of interaction rounds.
A round of communication consists of parallel message sending, 
reception, and local computation, with the constraint that 
messages for a given round are cannot depend on data from future rounds.

To facilitate the analysis, we assume all inputs are $n \times n$ square matrices 
with $\ell$-bit elements. The complexities of our protocols are 
summarized in Table \ref{tab:Complexity Analysis}.

\begin{table}[htbp]
  \centering
  \caption{Complexity Analysis of The PraxiMLP Framework}
    \resizebox{\linewidth}{!}{
    \begin{tabular}{lccc}
    \toprule
    \textbf{Secure Protocol} & 
    \textbf{Comp. Complexity} &
    \textbf{Comm. Overhead (bits)} &
    \textbf{Comm. Rounds} \\
    \midrule
    Shr($x$) & $\mathcal{O}(n^2)$ &  $3n^2\ell$ &   1 \\
    
    Rec($\llbracket x \rrbracket$) & $\mathcal{O}(n^2)$ &  
    $3n^2\ell$ &   1 \\

    $\llbracket x \rrbracket + \llbracket y \rrbracket$ & $\mathcal{O}(n^2)$ &  
    0 &   0 \\

    $\llbracket x \rrbracket \odot \llbracket y \rrbracket$ & $\mathcal{O}(n^2)$ &   
    $3n^2\ell$ &   1 \\

    $\llbracket x \rrbracket \times \llbracket y \rrbracket$ & $\mathcal{O}(n^3)$ & 
    $3n^2\ell$ &    1 \\

    Add2Mul($\llbracket x \rrbracket$) & $\mathcal{O}(n^2)$ &     
    $7n^2\ell$ &  2 \\
    
    Mul2Add($\llbracket x \rrbracket^M$) & $\mathcal{O}(n^2)$ & 
    $9n^2\ell$ &  2 \\

    ReLU($\llbracket x \rrbracket$) & $\mathcal{O}(n^2)$ & 
    $19n^2\ell$ &  5 \\

    Softmax($\llbracket x \rrbracket$) & $\mathcal{O}(n^2)$ & 
    $25n^2\ell$ &  6 \\

    \bottomrule
    \end{tabular}}
  \label{tab:Complexity Analysis}
\end{table}




\subsection{Practical Security Analysis}

The proposed protocols are provably secure in the practical 
semi-honest security model. A proof sketch is provided below:

\begin{proof}
    Security in our practical model requires that no party can infer 
    the exact value of another's private data from its view. This is achieved as follows: 
    the Rec($\llbracket x \rrbracket$) protocol is public by design, and the 
    addition protocol is inherently secure as it is purely local. The security of 
    all other protocols stems from masking private inputs with randomness 
    (e.g., using a random sharing of zero). This ensures that all protocol 
    messages appear statistically random to an adversary, thus revealing no precise 
    information about the underlying secrets. Therefore, all core protocols in the 
    PraxiMLP framework satisfy our definition of practical security.    
\end{proof}

Although participants cannot learn the precise value of a secret, 
they may be able to narrow its range based on prior knowledge. 
We illustrate this potential for range inference and its mitigation using 
the $\text{Shr}_0(x)$ protocol as an example.

Assume $P_0$ holds $x$ with a known prior range of $x \in [l_x, r_x]$. 
The Shr($x$) protocol masks $x$ using a zero sharing $(\alpha_0, \alpha_1, \alpha_2)$ 
to produce $\llbracket x \rrbracket = (x + \alpha_0, \alpha_1, \alpha_2)$. 
In this process, $P_1$ learns no information about $x$, while $P_2$ receives 
the share $x_0 = x + \alpha_0$. If $P_2$ also knows 
$\alpha_0 \in [l_\alpha, r_\alpha]$, it can update its belief 
about the range of $x$ to the intersection 
$[l_x,r_x] \cap [x_0 - r_\alpha, x_0 - l_\alpha]$.
Range inference occurs if this intersection is a tighter bound, 
when $x_0 - r_\alpha > l_x$ or $x_0 - l_\alpha < r_x$. 
This is prevented if $x_0$ falls within the "safe interval" 
$[r_x + l_\alpha, l_x + r_\alpha]$.
Assuming both $x$ and $\alpha_0$ are drawn from uniform distributions, 
let $L_x = r_x - l_x$, $L_\alpha = r_\alpha - l_\alpha$, 
and $\theta = L_\alpha / L_x$. 
The probability $Pr$ that the range of $x$ is not narrowed is:
\begin{align}
Pr = \frac{(l_x + r_\alpha) - (r_x + l_\alpha)}{(r_x + r_\alpha) - (l_x + l_\alpha)} 
= \frac{L_\alpha - L_x}{L_\alpha + L_x} = 1 - \frac{2}{\theta + 1}
\end{align}

Evidently, $Pr \to 1$ as $L_\alpha \gg L_x$. 
Since the zero-sharing is derived from a random sharing 
$\llbracket r \rrbracket = (r_0, r_1, r_2)$, assuming 
$r_i\in [l_r, r_r]$, the range of $\alpha_i$ becomes $[l_r - r_r, r_r - l_r]$, 
giving a range length of $L_\alpha = 2(r_r - l_r)$. 
Therefore, applications can enhance practical security by 
setting a sufficiently wide range of 
$\llbracket r \rrbracket$.

\section{Implementation and Evaluation}


The PraxiMLP framework is implemented in C++ using the YACL library 
for network communication, and is publicly (anonymously) available at 
\url{https://anonymous.4open.science/r/PraxiMLP}. 

\subsection{Experimental Setup}


All experiments were conducted on three servers 
with identical hardware configurations: 
22 vCPUs (Intel$^\circledR$ Core\texttrademark\ Ultra 7), 32 GB of RAM, 
running Ubuntu 22.04 LTS. The experiments were evaluated in two simulated 
network settings: a LAN setting 
(85 Gbps bandwidth, 0.04 ms Round-Trip Time) and a WAN setting 
(300 Mbps bandwidth, 40 ms Round-Trip Time).


We benchmark the PraxiMLP framework against four mainstream 
MPC-based PPML frameworks, including 
CrypTen, MP-SPDZ, SecretFlow, and SecureNN. 
Furthermore, the PyTorch framework is employed as the plaintext 
baseline for comparison in the MLP experiments.

\subsection{Basic Protocols}


We first evaluate the efficiency and precision of the four fundamental 
protocols required for the MLP model: matrix multiplication (MatMul), 
the matrix Hadamard product (MatHP), ReLU, and Softmax. 
All test data consists of randomly generated square matrices of 
floating-point numbers, with elements of the form 
$\pm a_0.a_1a_2\cdots a_{15}\times 10^\delta 
(\delta \in [-x,x], x\in \mathbb{Z})$. For the efficiency evaluation, 
we set $x=0$. In the precision evaluation, $x$ is increased from 0 to 8 in 
steps of 2 (i.e., $x \in \{0, 2, 4, 6, 8\}$).

\begin{figure}[htbp]
    \centering
    \subfigure[MatMul Comm. Overhead]
    {\includegraphics[width=0.485\linewidth]{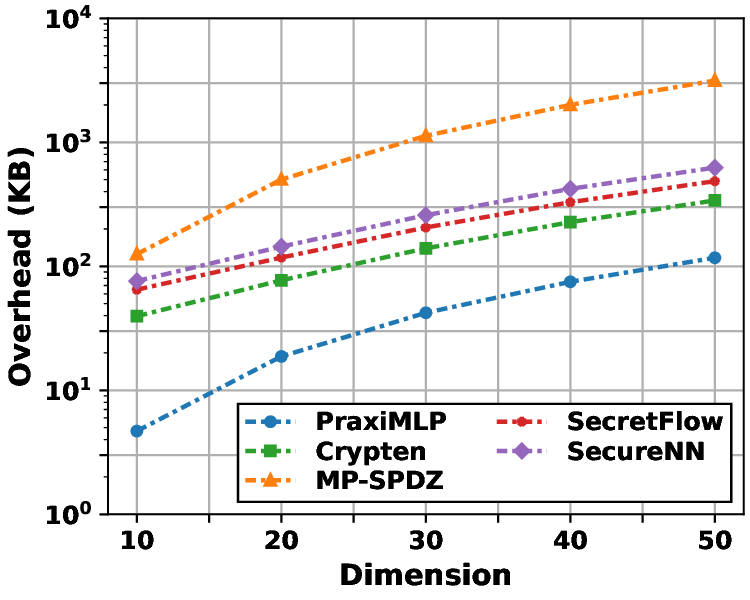}
    \label{fig: MatMul Comm.Overhead}}
    \subfigure[MatHP Comm. Overhead]
    {\includegraphics[width=0.485\linewidth]{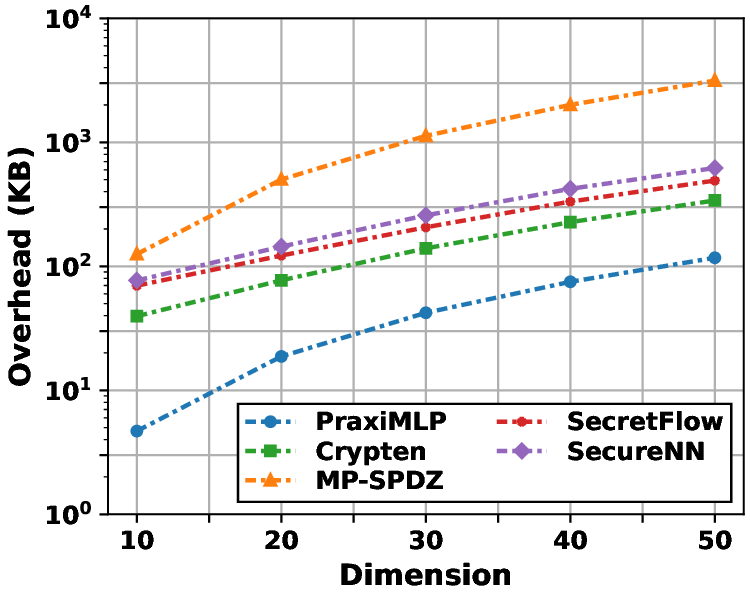}
    \label{fig: MatHP Comm.Overhead}}

    \centering
    \subfigure[ReLU Comm. Overhead]
    {\includegraphics[width=0.485\linewidth]{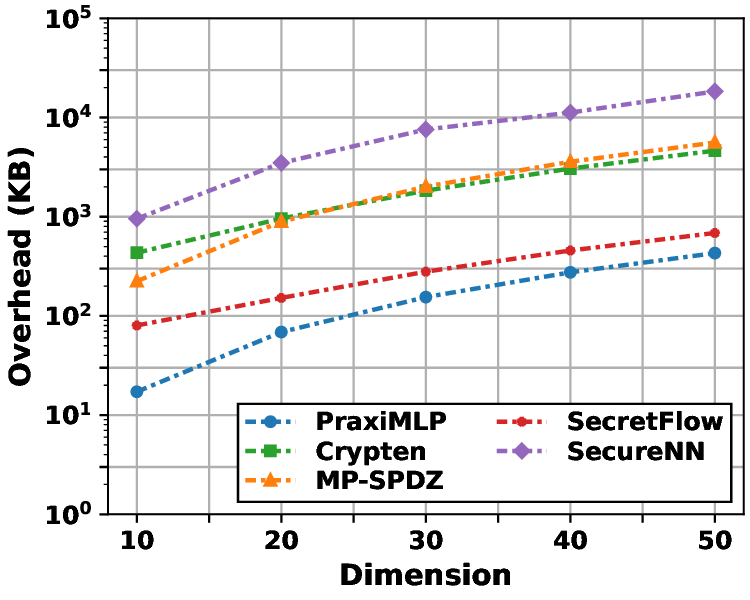}
    \label{fig: ReLU Comm.Overhead}}
    \subfigure[Softmax Comm. Overhead]
    {\includegraphics[width=0.485\linewidth]{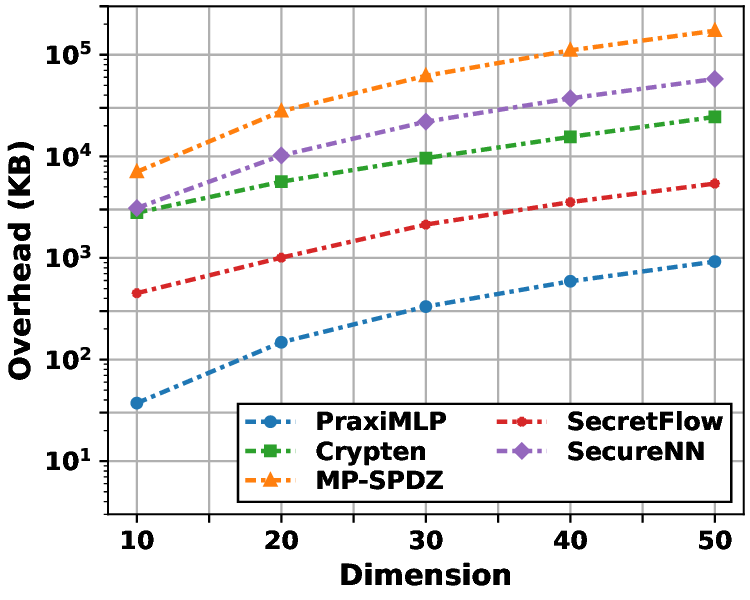}
    \label{fig: Softmax Comm.Overhead}}
    
    \caption{Communication Comparison of Basic Protocols}
    \label{fig: Communication Cost Comparison of Basic Protocols}
\end{figure}


\subsubsection{Efficiency Evaluation}
We benchmarked the running time and communication cost of each framework 
on the four fundamental protocols in the LAN setting, using $n \times n$ 
square matrices with $n \in \{10, 20, 30, 40, 50\}$ as input. 
All results are averaged over 1,000 runs. The running times are 
presented in Table \ref{tab: Runtime Comparison of Fundamental Protocols}, 
while the communication costs are plotted in 
Figure \ref{fig: Communication Cost Comparison of Basic Protocols}.



As shown in Table \ref{tab: Runtime Comparison of Fundamental Protocols} and Figure \ref{fig: Communication Cost Comparison of Basic Protocols}, the PraxiMLP framework significantly outperforms the other four frameworks in both runtime (achieving a maximum speedup of up to 19.9$\times$) and communication cost, which is substantially lower than its counterparts.

\begin{table}[h]
  \centering
  \caption{Runtime Comparison of Fundamental Protocols in the 
  LAN Environment, where Speedup indicates the speedup factor 
  of PraxiMLP relative to the second-best framework.}
  \resizebox{\linewidth}{!}{
    \begin{tabular}{ccrrrrrr}
    \toprule  
    \multicolumn{1}{c}{\multirow{2}[2]{*}{\textbf{Protocols}}} & 
    \multicolumn{1}{c}{\multirow{2}[2]{*}{\textbf{Dimension}}} & 
    \multicolumn{5}{c}{\textbf{Running Time (ms)}} & 
    \multicolumn{1}{c}{\multirow{2}[2]{*}{\textbf{Speedup}}}\\
    \cmidrule{3-7} 
    & & \textbf{CrypTen} & \textbf{MP-SPDZ} & \textbf{SecretFlow} & 
    \textbf{SecureNN} & \textbf{PraxiMLP}  \\
    \midrule
    \multirow{5}[2]{*}{\textbf{MatMul}} & 
    $10 \times 10$ & 5.78 & 2.35 & 48.41 & 49.77 & \textbf{0.30} & 7.9$\times$\\
    & $20 \times 20$ & 6.94 & 2.59 & 53.02 & 51.21 & \textbf{0.65} & 4.0$\times$\\
    & $30 \times 30$ & 7.19 & 4.79 & 55.55 & 52.59 & \textbf{0.67} & 7.1$\times$\\
    & $40 \times 40$ & 7.49 & 7.78 & 57.79 & 55.54 & \textbf{1.24} & 6.0$\times$\\
    & $50 \times 50$ & 8.54 & 11.48 & 59.97 & 59.56 & \textbf{1.41} & 6.0$\times$\\
    \midrule
    \multirow{5}[2]{*}{\textbf{MatHP}} & 
    $10 \times 10$ & 5.57 & 2.04 & 43.53 & 45.33 & \textbf{0.15} & 13.3$\times$\\
    & $20 \times 20$ & 6.08 & 3.72 & 46.10 & 47.67 & \textbf{0.26} & 14.2$\times$\\
    & $30 \times 30$ & 6.46 & 4.98 & 48.88 & 49.97 & \textbf{0.29} & 17.0$\times$\\
    & $40 \times 40$ & 7.16 & 6.10 & 50.89 & 51.42 & \textbf{0.31} & 19.9$\times$\\
    & $50 \times 50$ & 7.92 & 8.95 & 52.86 & 52.20 & \textbf{0.41} & 19.4$\times$\\
    \midrule
    \multirow{5}[2]{*}{\textbf{ReLU}} & 
    $10 \times 10$ & 19.51 & 1.73 & 38.82 & 48.89 & \textbf{1.48} & 1.2$\times$\\
    & $20 \times 20$ & 20.23 & 3.57 & 42.40 & 68.33 & \textbf{2.14} & 1.7$\times$\\
    & $30 \times 30$ & 25.08 & 5.55 & 46.33 & 102.15 & \textbf{2.55} & 2.2$\times$\\
    & $40 \times 40$ & 29.28 & 8.90 & 48.50 & 142.27 & \textbf{3.53} & 2.5$\times$\\
    & $50 \times 50$ & 32.30 & 13.46 & 49.40 & 194.46 & \textbf{5.09} & 2.6$\times$\\
    \midrule
    \multirow{5}[2]{*}{\textbf{Softmax}} & 
    $10 \times 10$ & 172.51 & 23.40 & 51.25 & 79.26 & \textbf{2.57} & 9.1$\times$\\
    & $20 \times 20$ & 191.92 & 55.66 & 60.31 & 124.60 & \textbf{3.89} & 14.3$\times$\\
    & $30 \times 30$ & 225.93 & 123.32 & 65.63 & 190.36 & \textbf{4.89} & 13.4$\times$\\
    & $40 \times 40$ & 233.18 & 189.92 & 72.61 & 295.30 & \textbf{8.05} & 9.0$\times$\\
    & $50 \times 50$ & 247.05 & 287.06 & 77.26 & 417.37 & \textbf{8.60} & 9.0$\times$\\
    \bottomrule
    \end{tabular}}
  \label{tab: Runtime Comparison of Fundamental Protocols}
\end{table}

The reason for this performance gap lies in the underlying 
protocol design. Frameworks such as CrypTen must typically 
convert arithmetic shares into Boolean shares or Yao's circuits 
to handle comparisons or non-linear operations (e.g., ReLU).
This cross-domain conversion process is known to incur significant 
communication overhead. In contrast, PraxiMLP leverages its novel 
conversion protocols between additive and multiplicative sharings to operate entirely 
within the arithmetic domain. This approach obviates the need for 
expensive share-type conversions and relies only on efficient matrix 
operations over arithmetic shares, thus achieving 
superior overall efficiency.

\begin{figure}[htb]
    \centering
    \subfigure[MRE of MatMul]
    {\includegraphics[width=0.485\linewidth]{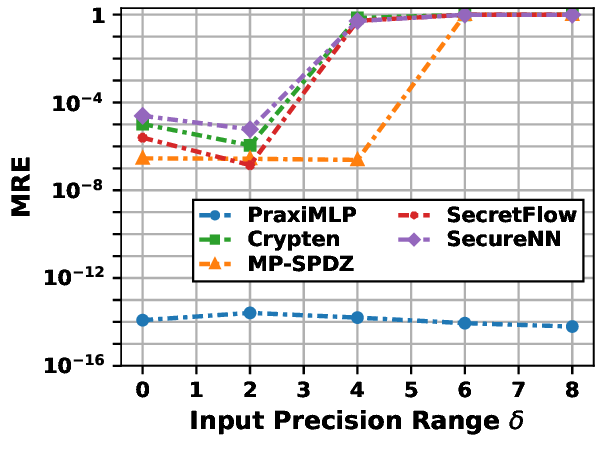}
    \label{fig: MRE of MatMul}}
    \subfigure[MRE of MatHP]
    {\includegraphics[width=0.485\linewidth]{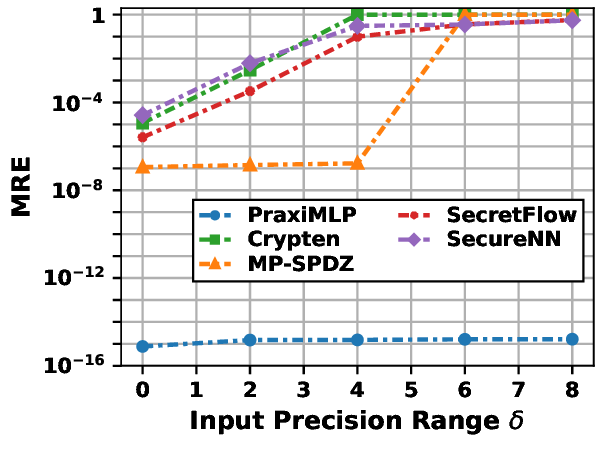}
    \label{fig: MRE of MatHP}}

    \centering
    \subfigure[MRE of ReLU]
    {\includegraphics[width=0.485\linewidth]{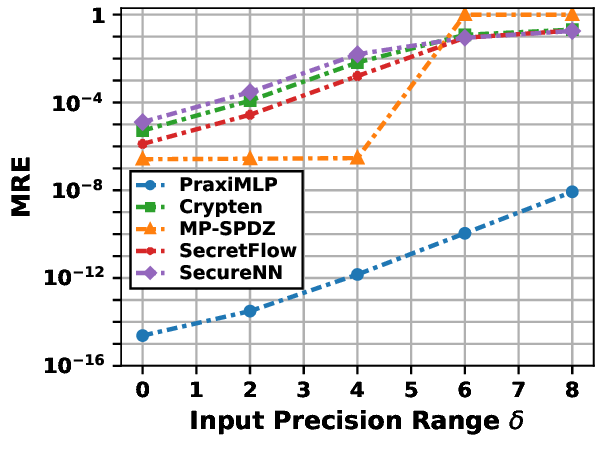}
    \label{fig: MRE of ReLU}}
    \subfigure[MRE of Softmax]
    {\includegraphics[width=0.485\linewidth]{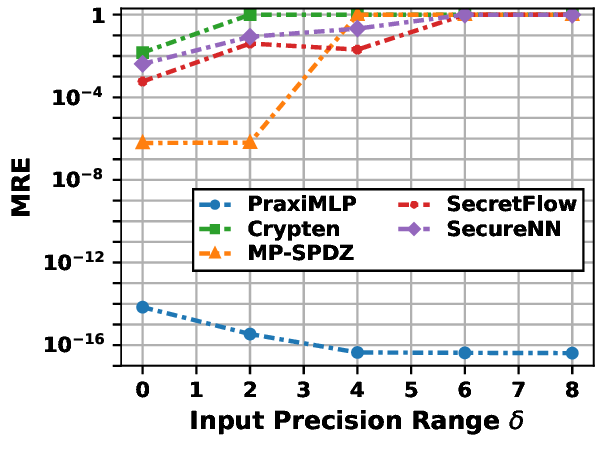}
    \label{fig: MRE of Softmax}}

    \caption{Precision Comparison of Fundamental Protocols}
    \label{fig: Precision Comparison of Fundamental Protocols}
\end{figure}

\subsubsection{Precision Evaluation}
The computational precision of each framework was then evaluated 
on the four fundamental protocols. This evaluation used $50 \times 50$ 
randomly generated floating-point matrices as input, 
with elements sampled from five distinct exponent ranges 
($\delta \in [-x,x]$, where $x \in \{0, 2, 4, 6, 8\}$). 
The Mean Relative Error (MRE) between the output of each 
framework and the plaintext result was calculated, averaged 
over 1,000 repetitions. These results are plotted in 
Figure \ref{fig: Precision Comparison of Fundamental Protocols}.


As shown in Figure \ref{fig: Precision Comparison of Fundamental Protocols}, 
PraxiMLP achieves dramatically superior precision, 
outperforming all counterparts by an average of 8 orders of magnitude. 
This advantage is rooted in our framework's native support for 
high-precision floating-point arithmetic, which avoids 
the significant error accumulation inherent in the fixed-point 
approximations used by other frameworks. 
This limitation is starkly highlighted when 
$\delta$ exceeds the $[-4, 4]$ range, 
where competing frameworks fail (MRE=1.0).
We also observe that the MRE for MatMul slightly decreases with 
larger $\delta$ due to a decreasing matrix condition number. 
Similarly, the Softmax error drops as the exponential function's 
output saturates, stabilizing the computation.

\subsection{MLP Model}

\subsubsection{Dataset and Model Architecture}
In the MLP model evaluation, four standard datasets were used: 
Iris\cite{cite-Iris}, Wine\cite{cite-Wine}, Mnist\cite{cite-Mnist}, 
and Fashion Mnist\cite{cite-Fashion-Mnist}, where Iris and Wine 
are 3-class classification datasets, 
whereas Mnist and Fashion Mnist are 10-class datasets.
The datasets were partitioned vertically (on the feature dimension) into three shares,
which were then distributed via PRSS such that each party holds two. For all experiments, 
the models were trained for five epochs. Further details on the 
datasets and model configurations are provided in Table \ref{tab:Datasets}.

\begin{table}[h]
  \centering
  \caption{Datasets and MLP Model Details}
    \resizebox{\linewidth}{!}{
    \begin{tabular}{lccccl}
    \toprule
    \textbf{Dataset} & 
    \textbf{Features} &
    \textbf{Train / Test} &
    \textbf{Batch Size} &
    \textbf{Learning Rate} &
    \textbf{Model}
    \\
    \midrule
   \textbf{Iris} & 4 &  120 / 30 & 16 & 0.05 & 16-16-3 \\
   \textbf{Wine} & 13 &  142 / 36 & 16 & 0.05 & 16-16-3 \\
   \textbf{Mnist} & 784 &  60,000 / 10,000 & 128 & 0.01 & 128-128-10 \\
   \textbf{Fashion} & 784 &  60,000 / 10,000 & 128 & 0.01 & 128-128-10 \\
    \bottomrule
    \end{tabular}}%
  \label{tab:Datasets}%
\end{table}%

\subsubsection{Efficiency and Accuracy Evaluation}
We evaluated the running time, communication cost, and accuracy of 
each framework for both MLP training and inference phases 
in both LAN and WAN settings. The experiments were repeated 50 times 
in the LAN setting and 5 times in the WAN setting, with the average values 
reported. The results are presented in 
Table \ref{tab: Performance Comparison of MLP}.

\begin{table}[htbp]
\centering
\caption{Performance Comparison of MLP on Four Datasets}
\label{tab: Performance Comparison of MLP}

\resizebox{\columnwidth}{!}{
\begin{tabular}{cl rrr rrr c}
\toprule
\multirow{2}[2]{*}{\textbf{Dataset}} & \multirow{2}[2]{*}{\textbf{Framework}} & 
\multicolumn{3}{c}{\textbf{Training}} & \multicolumn{3}{c}{\textbf{Predicting}} & 
\multirow{2}[2]{*}{\textbf{Accuracy}} \\
\cmidrule(lr){3-5} \cmidrule(lr){6-8}
 & & \textbf{Comm. (MB)} & \textbf{LAN (s)} & \textbf{WAN (s)} & \textbf{Comm. (MB)} & \textbf{LAN (s)} & \textbf{WAN (s)} & \\
\midrule
\multirow{6}{*}{\textbf{Iris}} 
 & CrypTen & 257 & 16.0 & 2105 & 2.2 & 0.22 & 27.0 & 86.67\% \\
 & MP-SPDZ & 296 & 2.1 & 234 & 4.0 & 0.29 & 1.8 & 93.33\% \\
 & SecretFlow & 23 & 4.0 & 168 & 0.3 & 0.07 & 1.3 & 93.33\% \\
 & SecureNN & 261 & 6.0 & 392 & 8.2 & 0.12 & 2.8 & 80.00\% \\
 & \textbf{PraxiMLP} & \textbf{5} & \textbf{0.2} & \textbf{30} & \textbf{0.2} & \textbf{0.004} & \textbf{0.6} & \textbf{96.67\%} \\
 \cmidrule(lr){2-9}
 & PyTorch & \textendash & 0.02 & \textendash & \textendash & 0.001 & \textendash & 100.00\% \\
\midrule
\multirow{6}{*}{\textbf{Wine}} 
 & CrypTen & 306 & 18.0 & 2374 & 2.6 & 0.23 & 27.1 & 97.22\% \\
 & MP-SPDZ & 356 & 2.1 & 264 & 4.9 & 0.25 & 2.5 & 94.44\% \\
 & SecretFlow & 28 & 4.6 & 189 & 0.4 & 0.08 & 1.3 & 97.22\% \\
 & SecureNN & 308 & 7.1 & 441 & 9.9 & 0.14 & 2.9 & 83.33\% \\
 & \textbf{PraxiMLP} & \textbf{7} & \textbf{0.2} & \textbf{34} & \textbf{0.3} & \textbf{0.004} & \textbf{0.6} & \textbf{100.00\%} \\
 \cmidrule(lr){2-9}
 & PyTorch & \textendash & 0.03 & \textendash & \textendash & 0.001 & \textendash & 100.00\% \\
\midrule
\multirow{6}{*}{\textbf{Mnist}} 
 & CrypTen & 306,285 & 16,973.9 & 144,773 & 1,074.6 & 17.96 & 889.0 & 92.17\% \\
 & MP-SPDZ & 1,517,429 & 3,986.6 & 660,133 & 10,221.2 & 36.54 & 2,491.0 & 96.65\% \\
 & SecretFlow & 69,300 & 519.1 & 13,869 & 1,376.0 & 10.95 & 143.7 & 96.79\% \\
 & SecureNN & 616,584 & 5,313.4 & 55,475 & 14,517.1 & 153.80 & 915.6 & 87.19\% \\
 & \textbf{PraxiMLP} & \textbf{25,077} & \textbf{306.4} & \textbf{2,371} & \textbf{505.1} & \textbf{5.35} & \textbf{56.8} & \textbf{96.98\%} \\
 \cmidrule(lr){2-9}
 & PyTorch & \textendash & 5.8 & \textendash & \textendash & 0.05 & \textendash & 97.19\% \\
\midrule
\multirow{6}{*}{\textbf{Fashion}} 
 & CrypTen & 285,019 & 16,922.5 & 144,569 & 1,007.8 & 19.21 & 888.7 & 84.16\% \\
 & MP-SPDZ & 1,504,992 & 4,080.7 & 619,887 & 10,069.5 & 36.90 & 2,499.0 & 85.79\% \\
 & SecretFlow & 69,566 & 497.2 & 13,695 & 1,366.1 & 11.25 & 144.3 & 85.62\% \\
 & SecureNN & 596,790 & 5,251.4 & 55,956 & 14,438.3 & 148.71 & 909.0 & 80.87\% \\
 & \textbf{PraxiMLP} & \textbf{25,077} & \textbf{328.3} & \textbf{2,359} & \textbf{505.1} & \textbf{5.56} & \textbf{56.7} & \textbf{86.18\%} \\
 \cmidrule(lr){2-9}
 & PyTorch & \textendash & 7.1 & \textendash & \textendash & 0.06 & \textendash & 87.06\% \\
\bottomrule
\end{tabular}%
} 
\end{table}


As shown in Table \ref{tab: Performance Comparison of MLP}, PraxiMLP 
achieves superior efficiency over all four counterparts while 
maintaining accuracy comparable to the plaintext baseline. 
The performance of the competing frameworks varies: 
SecretFlow, which utilizes an optimized ABY3 protocol, 
demonstrates strong computational efficiency. Although 
SecureNN avoids the use of Yao's circuits, its security 
guarantee against a single malicious party results in higher 
protocol complexity, which in turn negatively impacts its overall 
efficiency. The MP-SPDZ framework incurs significant communication 
overhead, degrading its performance, particularly on large-scale 
datasets and in the WAN setting. CrypTen, being implemented in Python 
and not utilizing replicated secret sharing techniques, exhibits the 
poorest efficiency.


In contrast, the PraxiMLP framework is designed to directly address these trade-offs. Grounded in the principle of practical security, its novel protocols operate entirely within the arithmetic domain. This design directly obviates the expensive cross-domain conversions, which is the primary source of its substantial boost in efficiency. Furthermore, PraxiMLP ensures high model accuracy by operating natively on high-precision floating-point numbers, bypassing the error accumulation inherent in the fixed-point approximations used by all competitors. These experiments demonstrate that the PraxiMLP framework provides an excellent balance of high efficiency and high precision, confirming its effectiveness for practical applications.

\section{Conclusion}


Grounded in the principle of practical security, this paper introduces 
novel conversion protocols between additive and multiplicative  
replicated secret sharing. Based on this foundation, it constructs PraxiMLP, 
a secure three-party MLP framework that operates entirely within 
the arithmetic sharing domain. Its native support 
for high-precision floating-point arithmetic enables precise 
implementations of the critical non-linear operations in MLP models. 
Experimental results demonstrate that PraxiMLP is significantly superior 
to most existing mainstream PPML frameworks in both efficiency and precision, 
on both fundamental protocols and end-to-end MLP models, thus validating 
the practicality and effectiveness of its design.

However, the practical security model adopted in this paper is, 
at present, more of a heuristic exploration. Future research directions 
include: (1) formally defining this security model and its corresponding security proof methodologies;
(2) rigorously determining the privacy 
leakage boundaries under this model; 
(3) extending the underlying protocols to provide security against malicious adversaries. 
Furthermore, extending the 
framework to support more complex models 
(such as CNNs, RNNs, and even Large Language Models), 
as well as evaluating 
its performance in settings with a larger number of parties, 
are also key priorities for future work.

\balance

\bibliographystyle{ACM-Reference-Format}
\bibliography{biblio}

\end{document}